\documentclass[prd, aps, superscriptaddress, twocolumn, preprintnumbers, floatfix, nofootinbib]{revtex4}
\usepackage[dvips]{graphicx}
\usepackage{epsf}
\usepackage{amsmath}
\usepackage{amssymb}

\usepackage{graphicx}
\usepackage{dcolumn}
\usepackage{bm}

\def\be{\begin{equation}}
\def\ee{\end{equation}}
\def\bea{\begin{eqnarray}}
\def\eea{\end{eqnarray}}

\usepackage{color}

\begin{document}


\title{Breaking of Spatial Diffeomorphism Invariance, Inflation and the Spectrum of Cosmological
Perturbations}
\author{L.L. Graef$^{1,2}$ and R. Brandenberger$^{}$}
\email{leilagraef@usp.br, rhb@physics.mcgill.ca}

\affiliation{Department of Physics, McGill University, 
Montr\'eal, QC, H3A 2T8, Canada \\
$^{2}$ Instituto de F\'isica, Universidade de S\~ao Paulo, 
Rua do Mat\~ao travessa R, 05508-090, S\~ao Paulo, SP, Brazil}


\begin{abstract}

Standard inflationary models yield a characteristic signature of a primordial power spectrum with 
a red tensor and scalar tilt. Nevertheless, Cannone et al \cite{Cannone} recently suggested that, 
by breaking the assumption of spatial diffeomorphism invariance in the context of the effective 
field theory of inflation, a blue tensor spectrum can be achieved without violating the Null Energy 
Condition. In this context, we explore in which cases a blue tensor tilt can be obtained along 
with a red tilt in the scalar spectrum. Ultimately, we analyze under which conditions this model 
can reproduce the specific consistency relation of String Gas Cosmology.
 
\end{abstract}

\maketitle

\section{Introduction}

One of the key predictions of the inflationary universe scenario \cite{Guth}
is that there should be a nearly scale-invariant spectrum of gravitational
waves with a {\it red tilt}, i.e. slightly more power on larger wavelengths
than on smaller ones \cite{Mukh}. This prediction stems from the basic setup of
inflation: coupling of a scalar matter field obeying the Null Energy
Condition to Einstein gravity. In this setup, the Hubble expansion
parameter $H$ is a slowly decreasing function of time during the period
of slow-roll inflation. The amplitude of the gravitational wave spectrum 
on a scale $k$ is set by the value of $H$ at the time when that mode
exits the Hubble radius $H^{-1}$. Hence, short wavelength modes which 
exit the Hubble radius at a later time obtain a smaller amplitude of the 
gravitational wave spectrum \cite{MFB, RHBfluctrev}. If $P_T(k)$ denotes the dimensionless
power spectrum of gravitational waves as a function of comoving wavenumber
$k$, and the index $n_T$ is defined by
\be
P(k) \, \sim \, k^{n_T}, \,
\ee
then the prediction of standard inflation is
\be \label{red}
n_T \, < \, 0 \, ,
\ee
i.e. a red tilt. In fact, in the case of single field slow-roll inflation with slow roll
parameter $\epsilon > 0$, the index is given by 
(see e.g. the review \cite{Liddle})
\be \label{cons1}
n_T \, = \, - 2 \epsilon \, .
\ee

However, inflation is not the only early universe scenario which is consistent
with the current data on cosmic microwave background (CMB)
anisotropies and large-scale structure (see e.g. \cite{RHBalt}
for a recent review of some alternatives). String Gas Cosmology
\cite{BV, NBV, SGCrev} is an alternative early universe scenario
which follows from merging basic principles of superstring
theory with ideas from cosmology. String Gas Cosmology predicts
a spectrum of cosmological perturbations with a blue tilt,
\be \label{basic}
n_T \, > \, 0 \, .
\ee
In fact, in the toy model of String Gas Cosmology developed in
\cite{BV, NBV}, there is a consistency relation between the
index $n_s$ of the scalar spectrum (which corresponds to
a red spectrum) and that of the tensor modes \cite{BNPV2, BNP}, which is given by
\be \label{cons2}
n_T \, \simeq \, - (n_s - 1) \, .
\ee

Current observations indicate that $n_s = 0.96 \pm 0.01$ \cite{Planck}.
Assuming that the ratio $r$ of the tensor spectrum to the scalar
spectrum is not too small ($r > 0.05$), then future CMB polarization 
measurements have the potential to differentiate between the single field
inflationary consistency relation (\ref{cons1}) and the String Gas
consistency relation (\ref{cons2}) \cite{Cora, Gabrielle}.

In the context of inflation, whereas the index $n_s$ of the scalar spectrum can be made
larger than $1$ for certain ranges of wavenumbers by introducing
more complicated scalar field actions, the tensor spectrum will
remain red, i.e. (\ref{red}) will remain true. The relation (\ref{red})
can be violated by abandoning the requirement that matter
satisfy the Null Energy Condition. An example is G-inflation
\cite{Ginfl}. However, the concern is that such models might not
be embeddable in an ultraviolet complete theory of matter
and gravity \cite{Adams}. 

Recently, the suggestion was made 
\cite{Cannone} that a blue tensor spectrum may result if one 
abandons some of the symmetries 
usually taken for granted in inflationary cosmology. In single field 
inflation the temporal diffeomorphism invariance is broken due to 
the time-dependence of the background.
It might very well be possible, however, that in the Lagrangean of 
fluctuations also the spatial diffeomorphism is broken. This possibility 
was considered in \cite{Cannone} in the context of the effective field 
theory of inflation.

The effective field theory approach corresponds to the description 
of a system through the lowest dimension
operators compatible with the underlying symmetries. This theory has 
been applied, in the last years, to describe the theory of fluctuations 
around an inflating cosmological background \cite{effectivetheory}. 
This approach allows us to characterize all the possible high
energy corrections to simple slow-roll inflation, whose sizes are 
constrained by experiment.  Also, it has the advantage of 
describing in a common language all single field models of inflation 
by using only symmetry principles.

It was recently shown \cite{Cannone} that the extra terms in the
effective field theory Lagrangian which arise if one allows for 
the breaking of spatial diffeomorphism invariance can produce 
a blue tilt of the tensor spectrum without assuming the presence
of matter which violates the Null Energy Condition. In this note 
we wish to explore whether it is possible to produce a blue tilt in 
the tensor spectrum (Section II) while maintaining a red tilt in the 
scalar spectrum (Section III). More specifically, we wish to determine if it is
possible to obtain the consistency relation  (\ref{cons2})
of String Gas Cosmology in this context. 
\\

\section{The Tensor Spectrum}

We are going to follow the approach of \cite{Cannone} and consider the 
effective field theory for cosmological perturbations around a de Sitter 
background, with temporal and spatial diffeomorphism invariance broken. Despite the breaking of spatial diffeomorphism invariance, in this scenario isotropy is preserved. The case in which anisotropies can be generated was considered in the recent paper \cite{Cannone2}.
  
The procedure usually adopted in the effective theory considers that 
the scalar mode can be eaten by the metric by going to unitary gauge. In this 
case there is no perturbation of the inflaton field and all the three degrees 
of freedom \footnote{Note that here we are not counting the vector modes.}
are in the metric (the scalar mode and the two tensor helicities). 
This setup is analogous to what happens in a spontaneously broken gauge 
theory. For simplicity we will focus on operators at most quadratic in the 
fluctuations.

We start with the background metric
\begin{equation}
ds^{2} \, = \, \bar{g}_{\mu\nu}(\eta) dx^{\mu} dx^{\nu} \, = \, a^{2}(\eta)(-\eta_{\mu\nu} dx^{\mu} dx^{\nu}),
\end{equation}
where $\eta$ is conformal time, the $x_i$ are comoving spatial
coordinates, $a^{2}(\eta)$ is the scale factor and $a(\eta)=1/(-H\eta)$ for a
background de Sitter space. The perturbed metric is
\begin{equation}
ds^{2} \, = \, g_{\mu\nu}(x, \eta) dx^{\mu} dx^{\nu}, 
\end{equation}
and the metric fluctuations are defined by
\be
h_{\mu\nu}(x, \eta) \, = \, g_{\mu\nu}(x, \eta) - \bar{g}_{\mu\nu}(\eta)
\ee
and are assumed to be small in amplitude (compared to the 
background metric).

The breaking of spatial diffeomorphism invariance will be described
through effective mass terms in the action for cosmological perturbations.
These terms do not necessarily
originate from a theory of massive gravity but simply correspond to the
most general way to express quadratic non-derivative operators in the 
fluctuations that break the spatial diffeomorphism symmetry.

Thus, to the  usual Einstein-Hilbert action expanded to second order, 
we add generic operators with no derivatives that are quadratic in 
the metric fluctuations $h_{\mu\nu}$,
\begin{align} \label{pdf1}
& S= \int d^{4}x \sqrt{-g} M_{pl}^{2} \left[R - 2\Lambda -2cg^{00}\right] \nonumber\\
& +\frac{1}{4} M_{pl}^{2} \int d^{4}x \sqrt{-g} [m_{0}^{2} h_{00}^{2} + 2m_{1}^{2} h_{0i}^{2} - m_{2}^{2} h_{ij}^{2} \nonumber\\ 
& + m_{3}^{2} h_{ii}^{2} - 2m_{4}^{2} h_{00} h_{ii}]. 
\end{align}
The terms in the first line are the only ones which contribute to the 
homogeneous and isotropic background. The terms linear in the
metric fluctuations from the first line vanish if the background is a solution
of the equations of motion. The terms quadratic in $h_{\mu \nu}$ from
the first line are those which arise in the usual theory of
cosmological perturbations. The new term proportional to 
$m_{0}^{2}$ breaks time reparametrization invariance and the 
other mass terms break invariance under spatial diffeomorphisms. In the limit $m_{i} \rightarrow 0$ with 
$i \neq 0$ the invariance is restored. 

We can consider the mass terms in the above equation as arising  
from couplings between the metric and fields with a 
time-dependent profile during inflation. As an approximation, we assume that
their coefficients are effectively constant
in space and time during inflation, while they go to zero after inflation ends. 
However, a small time-dependence proportional to the slow-roll parameter 
should be expected for these coefficients.

We can write equation (\ref{pdf1}) in terms of scalar, vector and tensor 
perturbations, by decomposing the fluctuations as follows,
\begin{align}
& h_{00} =\psi, \\
\nonumber & h_{0i} = u_{i} + \partial_{i} v,   \; \;  \; \; \; \; \; \; \; \; \; \; \; \; \; \; \; \; \; \; \; \; \; \; \; \; \; \; \; \; with \; \;   \partial_{i} u_{i} =0, \\
\nonumber & h_{ij} = \chi_{ij} + \partial_{(iSj)} + \partial_{i} \partial_{j} \sigma + \delta_{ij} \tau, \; \; with \; \partial_{iSi} =\partial_{i} \chi_{ij}=0. 
\end{align}

From the tensor part of the action, Cannone et al. \cite{Cannone} obtained 
the following spectrum
\bea \label{pdftensorspectrum}
P_{T} \, &=& \, \frac{2H^{2}}{\pi^{2} M_{pl}^{2} c_{T}} \left(\frac{k}{k_{*}}\right)^{n_{T}}, \nonumber \\
n_{T} \, &=&-\, 2\epsilon + \frac{2}{3} \frac{m_{2}^{2}}{H^{2}} \left(1+ \frac{4}{3}\epsilon\right),
\eea
to first order in the slow-roll parameter. In the above equation 
$c_{T}=1$ if we consider only the mass terms previously described, 
neglecting possible higher derivative terms in the Lagrangean. Note
that the parameters $m_0$, $m_1$, $m_3$ and $m_4$ do not
appear in the action for tensor perturbations.

We can see from the above equation that, if $m_{2}^{2}/H^{2}$ is 
positive and sufficiently larger than the slow-roll parameter, then
we obtain a positive tensor spectral index \cite{Cannone}. 
This is an interesting result since it shows that a blue tensor spectrum
like in String Gas Cosmology can be obtained in a modified
inflationary setup without violating the Null Energy Condition. 

\section{The Scalar Spectrum}

Naively, we could fear that the model considered here would also
give a blue scalar spectrum, and hence be in contradiction
with current  observational constraints. We will now show that
this is not the case, the reason being that the action contains
more free parameters than simply $m_2$. We will now
derive the constraints on the 
parameters $m_{i}$ from the scalar spectrum.

Expanding the action (\ref{pdf1}) to second order in the scalar fluctuations 
and substituting in it the equation of motion of the auxiliary fields 
$\psi$, $v$ and $\sigma$, we obtain after some algebra 
\begin{align} \label{pdf24}
&S= M_{pl}^{2} \int d^{4}x \frac{a^{2}}{H^{2}} [\frac{(m_{0}^{2} + 2 \epsilon H^{2})(m_{2}^{2} - m_{3}^{2})+m_{4}^{4}}{2(m_{2}^{2} - m_{3}^{2})} \tau'^{2} \nonumber \\ & + \epsilon H^{2} \tau \nabla^{2} \tau - \frac{m_{2}^{2} a^{2} H^{2} (m_{2}^{2} - 3m_{3}^{2} + (3+ \epsilon)m_{4}^{2})}{m_{2}^{2} - m_{3}^{2}} \tau^{2}],
\end{align}
which is a function of a single field $\tau$.
In deriving the above equation, the parameter $m_{1}^{2}$ was 
chosen to be zero in order to eliminate degrees of freedom, 
since in this case it can be shown that no vector modes propagate.

The scalar perturbation $\tau$ is related to the comoving curvature 
perturbation through the equation
\begin{equation}
\mathcal{R} \, = \, \tau - \frac{\mathcal{H}(\tau' - \mathcal{H} \psi)}{\mathcal{H}' - \mathcal{H}^{2}} \, .
\end{equation}
In unitary gauge, the equation of motion of the auxiliary scalar field 
$\psi$ leads to $\tau'=\mathcal{H} \psi$, and therefore we have 
$\mathcal{R}=\tau$. It is possible to show that  in this model
the curvature perturbation is not constant on super-Hubble scales. 
Moreover, the comoving curvature perturbation  $\mathcal{R}$ 
and the curvature perturbation on uniform density slices  $\zeta$ 
do not coincide in the large scale limit, unlike what happens if
spatial diffeomorphism invariance is unbroken. 

We can normalize $\tau$ by substituting $\hat{\tau}^{2}= N^{2}\tau^{2}$ 
in equation (\ref{pdf24}), $N^{2}$ being defined as
\begin{equation} \label{pdfN}
N^{2} \, \equiv \, \left(\frac{M_{pl}^{2}}{H^{2}}\right) \frac{(m_{0}^{2} + 2 \epsilon H^{2})(m_{2}^{2} - m_{3}^{2})+m_{4}^{4}}{2(m_{2}^{2} - m_{3}^{2})} \, .
\end{equation}
Then we can write the action in the simpler form
\begin{equation} \label{pdf25}
S \, = \, \int d^{4} x a^{2} [\hat{\tau}'^{2} + c_{s}^{2}(\hat{\tau} \nabla^{2} \tau) + a^{2}M^{2}\hat{\tau}^{2}] \, ,
\end{equation}
where
\begin{equation} \label{pdfcs}
c_{s}^{2} \, = \, 2 \epsilon H^{2}\frac{(m_{2}^{2} - 
m_{3}^{2})}{(m_{0}^{2}+2\epsilon H^{2})(m_{2}^{2}-m_{3}^{2})+ m_{4}^{4}} \, ,
\end{equation}
(note that in the limit when the extra terms in the action go to zero, $c_s^2$ goes to $1$)
and
\begin{equation} \label{pdfM}
M^{2} \, = \, \frac{-2m_{2}^{2} H^{2}(m_{2}^{2}-3m_{3}^{2}+(3+\epsilon)m_{4}^{2})}{(m_{0}^{2}+2\epsilon H^{2})(m_{2}^{2}-m_{3}^{2})+ m_{4}^{4}} \, ,
\end{equation}
which vanishes in the limit that the coefficients leading to the breaking of the spatial diffeomorphism
invariance go to zero.

We can define a new variable $v \equiv a \hat{\tau}$ and write the action (\ref{pdf25}) 
in terms of this variable. Note that in the limit in which the coefficients which lead to the breaking
of spatial diffeomorphism invariance go to zero, we find that $v$
coincides with the standard Sasaki-Mukhanov variable \cite{Sasaki},
in terms of which the action has canonical form. In our case,
it follows from the Euler-Lagrange equation that 
\begin{equation}
v"  + c_{s}^{2}k^{2} v - [(1+2\epsilon)b +(2+3\epsilon)]\frac{v}{\eta^{2}} \, = \, 0 \, ,
\end{equation}
where $b \equiv M^{2}/H^{2}$. We have considered $a"/a \approx (2+3\epsilon)/\eta^{2}$, 
since $a\approx-(1+\epsilon)/H\eta$ during inflation. It is the presence of
$b \neq 0$ which leads to the fact that the comoving curvature fluctuation
variable is not constant on super-Hubble scales.

Assuming the  Bunch-Davis vacuum as initial condition, we have in the limit of small 
wavelengths the solution 
\be
v \, = \, e^{-ic_{s}k\eta}/\sqrt{2c_{s}k} \, .
\ee 
While in the large wavelength limit we have the solution
\begin{equation}
v \, = \, c_{1} \eta^{\frac{1}{2} - \alpha} + c_{2} \eta^{\frac{1}{2}+\alpha} \, ,
\end{equation}  
where we defined 
\be 
\alpha \, \equiv \, \sqrt{2\epsilon b+3\epsilon +b +9/4} \, .
\ee
The first term is the growing mode (supposing $b+9/4 \geq 0$). We are going to consider 
only this growing mode. 

If we match the solution for small and for large wavelengths when 
$c_{s}^{2}k^{2} = (b+2)/\eta^{2}$ \cite{elisa, yfu}, we obtain for the 
constant $c_{1}$ the following expression 
\begin{equation}
c_{1} \, = \, \frac{e^{-i\sqrt{b+2}}}{\sqrt{2c_{s} k}} \left(\frac{c_{s} k}{\sqrt{b+2}}\right)^{\frac{1}{2} - \alpha} \, .
\end{equation}

From these solutions, we obtain the following scalar power spectrum
\begin{equation} \label{pdfscalarspectrum}
P_{\mathcal{R}} \, = \,  \frac{k^{3}|v|^{2}}{a^{2}N^{2}} = k^{3-2\alpha}\left(\frac{c_{s}^{-2\alpha}}{2a^{2}N^{2}}\right)\left(\frac{\eta}{\sqrt{b+2}}\right)^{1-2\alpha} \, .
\end{equation}
Since  in this model the curvature perturbation is not constant after Hubble radius crossing, 
an explicit time-dependence appears in the above expression. We need to evaluate
the result at the end of inflation ($\eta= \rm{cst}$, $a = \rm{cst}$).

We can see that the scalar spectral index is given by
\begin{equation} \label{pdfns1}
n_{s} - 1\, = \, 3 -2\alpha \, = \, 3 - 2\sqrt{2\epsilon b + 3\epsilon +b + 9/4} \, .
\end{equation}
Thus, the spectrum will be red, $n_{s}<1$, if 
\be
\sqrt{2\epsilon b + 3\epsilon + b +9/4} \, > \, 3/2 \, .
\ee
We can see that, in order to obtain the observed value for the spectral index, 
$n_{s} = 0.96$, the parameter $b$ must be close to zero. This can occur if the 
condition $m_{2}^{2} \approx 3m_{3}^{2}$ is satisfied, in addition to the 
condition $m_{4} \approx 0$, (see eq. (\ref{pdfM})). Also we can see from 
equation (\ref{pdfscalarspectrum}), which is calculated at a fixed time 
(the end of inflation), that the expected amplitude can also be obtained if
$c_{s} \rightarrow 1$ and $N^{2} \rightarrow \epsilon$. One possibility to 
obtain these limits is if, in addition to  $m_{4} \approx 0$, the parameter 
$m_{0}$ is much smaller than $\sqrt{\epsilon} H$. Although this is not the 
only possibility that provides the observed value of the scalar spectrum, it
is an interesting specific case to be considered. We can see that none of 
these bounds implies in an upper limit for the parameter $m_{2}$. This proves 
that its possible in this model to have a blue tensor spectrum and at the same
time maintain the  observed red scalar spectrum. 

Comparing equations (\ref{pdfscalarspectrum}) and (\ref{pdfns1}) with the 
corresponding equations in the usual models of inflation, we can see that 
the slow-roll parameter that appears in the term $3\epsilon$ in the above 
equations must correspond to the slow-roll parameter calculated at the moment 
of Hubble radius exit. When $b \rightarrow 0$, the curvature perturbation is 
conserved after Hubble radius crossing and thus, in this limit, the slow-roll 
parameter is calculated at the Hubble radius exit. Therefore, we are going to 
denote this quantity by $\epsilon_{c}$.

We can write a simplified expression for  $n_{s} -1$ by expanding the square root 
in equation (\ref{pdfns1}) as follows,
\begin{equation} \label{pdfns2}
n_{s} -1 \, = \, 3 - 3\sqrt{\frac{8}{9} \epsilon b + \frac{4}{3} \epsilon_{c} 
+ \frac{4}{9}b +1} \, \approx \, - 2\epsilon_{c} - \frac{2}{3}b \, .
\end{equation}
This quantity must be approximately $-0.04$ according to observations. Since the usual inflationary models give good agreement with observations 
for this index, the above expression must correspond to 
$n_{s}-1 \approx -2\epsilon_{V}$, where $\epsilon_{V}$ is the slow-roll 
parameter in the usual inflationary models. Thus, we have the following relation,
\begin{equation} \label{pdfepsilons}
\epsilon_{c} +\frac{b}{3} \, \approx \, \epsilon_{V} \, .
\end{equation}
From this relation we can see that in the model considered here the value 
of the slow-roll parameter at the time of Hubble radius crossing can be
smaller (or bigger) than the usual one by an amount of $\approx b/3$.


Using equations (\ref{pdftensorspectrum}) and (\ref{pdfscalarspectrum}), we can 
calculate the tensor-to-scalar ratio defined as 
\be
r \, \equiv \, P_{t}(k_{*})/P_{S}(k_{*}) \, = \, A_{t}/A_{S} \, .
\ee
We then obtain
\begin{equation}
r \, = \,  \frac{2H^{2}}{\pi^{2}M_{pl}^{2}} 
\left(\frac{2a^{2}N^{2}}{c_{s}^{-2\alpha}}\right)\left(\frac{\sqrt{b+2}}{\eta}\right)^{1-2\alpha} \, ,
\end{equation}
where $N$ and $c_{s}$ are given by eqs. (\ref{pdfN}) and (\ref{pdfcs})  respectively. 
It is possible to verify that in the limit  $b=0$, $c_{s}=1$ and $N^{2}=\epsilon$, the 
expected expression for the tensor-to-scalar ratio is recovered. In this case, the 
usual consistency relation, $r = -8n_{t}$, is also recovered.

By comparing the expressions (\ref{pdftensorspectrum}) and (\ref{pdfns2}) for the 
tensor and scalar spectral index of the model considered, 
\begin{align}
\label{pdfntcomp} n_{T} \, = \,  -2\epsilon + \frac{2}{3} \frac{m_{2}^{2}}{H^{2}}(1+2\epsilon), 
\\ \label{pdfnscomp}
n_{s} -1 \, = \, -2\epsilon - \frac{2}{3} \frac{M^{2}}{H^{2}}(1+2\epsilon),
\end{align}
it is  possible to see that in the case the tensor spectrum is blue, when the second 
term in the expression (\ref{pdfntcomp}) is bigger than the first, one can obtain 
the String Gas Cosmology relation, $n_{t} \approx -(n_{s}-1)$. This relation is satisfied 
whenever
\begin{equation}
-2\epsilon + \frac{2}{3} \frac{m_{2}^{2}}{H^{2}}(1+2\epsilon) \, = \, +2\epsilon + \frac{2}{3} \frac{M^{2}}{H^{2}}(1+2\epsilon) \, .
\end{equation}
We should point out that $M^{2}$ can be positive or negative, but the right 
hand side of equation (\ref{pdfnscomp}) must be negative in order to provide 
a red scalar spectrum. 

The above equality can be satisfied for a scalar spectrum compatible with  
observations with a complementary blue tensor spectrum. Therefore, we 
conclude that the model considered here can reproduce the characteristic 
consistency relation of the String Gas Cosmology. Obviously, this requires
fine tuning of the parameters of the model.

The results introduced here have been obtained from mass operators 
in the Lagrangean that break spatial diffeomorphism invariance. Nevertheless, in 
\cite{Cannone} it was shown that certain operators containing more than two spatial 
derivatives can mimic the effects of these mass operators even in scenarios which 
preserve spatial diffeomorphism invariance.

\section{Discussion}

A primordial tensor power spectrum with a blue spectral tilt is a characteristic 
signature of some alternative models to inflation, in particular String Gas Cosmology,
but also of some inflationary models which violate the Null Energy Condition.
A possible detection of a blue tensor tilt would immediately falsify 
standard inflation (based on the usual symmetries and on assuming matter which 
satisfies the Null Energy Condition). Since from the point of view of an ultraviolet
complete theory of matter it is problematic to violate the Null Energy Condition
\cite{Adams}, it becomes important to carefully investigate different ways of producing
a blue tensor tilt.  

In this spirit, it has recently been pointed out \cite{Cannone} that the breaking of 
spatial diffeomorphism invariance during inflation could provide a new mechanism 
to generate a blue tensor tilt, without postulating matter which violates the 
Null Energy Condition. On the other hand, its well known that the tilt of the scalar spectrum is
observed to be red. Hence, in this paper we have studied under which conditions one
can obtain both a blue tensor tilt and a red scalar tilt compatible with current
observations. We have seen that this is possible with appropriate choices
of the new parameters in the effective Lagrangean. 

Assuming that the tensor-to-scalar ratio is not too small ($r>0.05$), then future 
CMB polarization measurements have the potential to differentiate the single 
field inflationary consistency relation, $n_{T} = 2\epsilon$, and the String Gas 
consistency relation, $n_{T} \simeq -(n_{s}-1)$ \cite{Cora, Gabrielle}. 
Motivated by this prospect, we
have studied whether it is possible for the
models introduced in \cite{Cannone} to produce scalar and
tensor tilts which agree with the consistency relation from String Gas Cosmology.
We have seen that, given special choices of the parameters, this
is also possible. 

Thus, we suggest that its important to investigate other 
predictions of this scenario, like non-Gaussianities, in order to find which observables 
could be able to distinguish the special class of models of \cite{Cannone}
which are consistent with the string gas consistency relation from the
actual String Gas Cosmology scenario. String Gas Cosmology predicts
negligible non-Gaussianities on cosmological scales \cite{SGNG},
whereas usual inflationary models predict typically small but non-negligible
non-Gaussianities. A similar proposal to distinguish various scenarios
consistent with a blue tensor tilt has been made in \cite{YiWang}. 

\acknowledgements{The authors are grateful to E.G.M. Ferreira and D. Cannone 
for the useful discussions. RB is supported by an NSERC Discovery Grant, and by 
funds from the Canada Research Chair program. LG is supported by FAPESP 
under grants 2012/09380-8 .}


\begin{thebibliography}{99}

\bibitem{Cannone}
D.~Cannone, G.~Tasinato and D.~Wands,
  ``Generalised tensor fluctuations and inflation,''
  JCAP {\bf 1501}, no. 01, 029 (2015)
  [arXiv:1409.6568 [astro-ph.CO]].

\bibitem{Guth}
A. Guth, 
``The Inflationary Universe: A Possible Solution To The Horizon And Flatness Problems,'' 
Phys.\ Rev.\  D {\bf 23}, 347 (1981);\\
K.~Sato, 
``First Order Phase Transition Of A Vacuum And Expansion Of The Universe,'' 
Mon.\ Not.\ Roy.\ Astron.\ Soc.\  {\bf 195}, 467 (1981);\\
A.~A.~Starobinsky, 
``A New Type of Isotropic Cosmological Models Without Singularity,'' 
Phys.\ Lett.\ B {\bf 91}, 99 (1980);\\
R.~Brout, F.~Englert and E.~Gunzig, 
``The Creation Of The Universe As A Quantum Phenomenon,'' 
Annals Phys.\  {\bf 115}, 78 (1978).

\bibitem{Mukh}
 V. Mukhanov and G. Chibisov, 
 ``Quantum Fluctuation And Nonsingular Universe. (In Russian),'' 
 JETP Lett.\  {\bf 33}, 532 (1981) [Pisma Zh.\ Eksp.\ Teor.\ Fiz.\  {\bf 33}, 549 (1981)].
  
\bibitem{MFB} V.~F.~Mukhanov, H.~A.~Feldman and R.~H.~Brandenberger, 
``Theory of cosmological perturbations. Part 1. Classical perturbations. Part 2. Quantum theory of perturbations. Part 3. Extensions,'' 
Phys.\ Rept.\  {\bf 215}, 203 (1992).
  
\bibitem{RHBfluctrev} R.~H.~Brandenberger, 
``Lectures on the theory of cosmological perturbations,''
 Lect.\ Notes Phys.\  {\bf 646}, 127 (2004) [arXiv:hep-th/0306071].
	
\bibitem{Liddle}
 A.~R.~Liddle and D.~H.~Lyth,
  ``The Cold dark matter density perturbation,''
  Phys.\ Rept.\  {\bf 231}, 1 (1993)
  [astro-ph/9303019].
  
\bibitem{RHBalt}
 R.~H.~Brandenberger,
  ``Cosmology of the Very Early Universe,''
  AIP Conf.\ Proc.\  {\bf 1268}, 3 (2010)
  [arXiv:1003.1745 [hep-th]];\\
R.~H.~Brandenberger,
  ``Unconventional Cosmology,''
  Lect.\ Notes Phys.\  {\bf 863}, 333 (2013)
  [arXiv:1203.6698 [astro-ph.CO]].
  
 \bibitem{BV}
R.~H.~Brandenberger and C.~Vafa, 
  ``Superstrings In The Early Universe,'' 
  Nucl.\ Phys.\ B {\bf 316}, 391 (1989).
 
\bibitem{NBV}
A.~Nayeri, R.~H.~Brandenberger and C.~Vafa, 
  ``Producing a scale-invariant spectrum of perturbations in a Hagedorn phase 
  of string cosmology,''
  Phys.\ Rev.\ Lett.\  {\bf 97}, 021302 (2006)
  [arXiv:hep-th/0511140];\\
 R.~H.~Brandenberger, A.~Nayeri, S.~P.~Patil and C.~Vafa,
  ``String gas cosmology and structure formation,''
  Int.\ J.\ Mod.\ Phys.\ A {\bf 22}, 3621 (2007)
  [hep-th/0608121].

\bibitem{SGCrev}
R.~H.~Brandenberger, A.~Nayeri, S.~P.~Patil and C.~Vafa,
  ``String gas cosmology and structure formation,''
  Int.\ J.\ Mod.\ Phys.\ A {\bf 22}, 3621 (2007)
  [hep-th/0608121];\\
  T.~Battefeld and S.~Watson,
  ``String gas cosmology,''
  Rev.\ Mod.\ Phys.\  {\bf 78}, 435 (2006)
  [hep-th/0510022];\\
R.~H.~Brandenberger,
  ``String Gas Cosmology,''
  arXiv:0808.0746 [hep-th]\\
 R.~H.~Brandenberger,
  ``String Gas Cosmology: Progress and Problems,''
  Class.\ Quant.\ Grav.\  {\bf 28}, 204005 (2011)
  [arXiv:1105.3247 [hep-th]].
  
\bibitem{BNPV2}
R.~H.~Brandenberger, A.~Nayeri, S.~P.~Patil and C.~Vafa,
  ``Tensor Modes from a Primordial Hagedorn Phase of String Cosmology,''
  Phys.\ Rev.\ Lett.\  {\bf 98}, 231302 (2007)
  [hep-th/0604126].

 \bibitem{BNP}
 R.~H.~Brandenberger, A.~Nayeri and S.~P.~Patil,
  ``Closed String Thermodynamics and a Blue Tensor Spectrum,''
  Phys.\ Rev.\ D {\bf 90}, no. 6, 067301 (2014)
  [arXiv:1403.4927 [astro-ph.CO]].

\bibitem{Planck}
P.~A.~R.~Ade {\it et al.}  [Planck Collaboration],
  ``Planck 2015 results. XIII. Cosmological parameters,''
  arXiv:1502.01589 [astro-ph.CO];\\
P.~A.~R.~Ade {\it et al.}  [BICEP2 and Planck Collaborations],
  ``Joint Analysis of BICEP2/$Keck  Array$ and $Planck$ Data,''
  Phys.\ Rev.\ Lett.\  {\bf 114}, no. 10, 101301 (2015)
  [arXiv:1502.00612 [astro-ph.CO]].
  
\bibitem{Cora}
 L.~Boyle, K.~M.~Smith, C.~Dvorkin and N.~Turok,
  ``On testing and extending the inflationary consistency relation for tensor modes,''
  arXiv:1408.3129 [astro-ph.CO].
  
\bibitem{Gabrielle}
G.~Simard, D.~Hanson and G.~Holder,
  ``Prospects for Delensing the Cosmic Microwave Background for Studying Inflation,''
  arXiv:1410.0691 [astro-ph.CO].
  
\bibitem{Ginfl}
T.~Kobayashi, M.~Yamaguchi and J.~Yokoyama,
  ``G-inflation: Inflation driven by the Galileon field,''
  Phys.\ Rev.\ Lett.\  {\bf 105}, 231302 (2010)
  [arXiv:1008.0603 [hep-th]].
  
\bibitem{Adams}
A.~Adams, N.~Arkani-Hamed, S.~Dubovsky, A.~Nicolis and R.~Rattazzi,
  ``Causality, analyticity and an IR obstruction to UV completion,''
  JHEP {\bf 0610}, 014 (2006)
  [hep-th/0602178].
	
\bibitem{effectivetheory} C. Cheung, P. Creminelli, A. L. Fitzpatrick, J. Kaplan, L. Senatore,
``The Effective Field Theory of Inflation,''
JHEP {\bf 0803}, 014 (2008) [hep-th/0709.0293].	

\bibitem{Cannone2} D. Cannone, J. Gong, G. Tasinato, `` Breaking discrete symmetries in the effective field theory of inflation,'' arXiv:1505.05773 [hep-th].

\bibitem{Sasaki}
M.~Sasaki, 
``Large Scale Quantum Fluctuations in the Inflationary Universe,'' 
Prog.\ Theor.\ Phys.\  {\bf 76}, 1036 (1986).;\\
V.~F.~Mukhanov, 
``Quantum Theory of Gauge Invariant Cosmological Perturbations,'' 
Sov.\ Phys.\ JETP {\bf 67}, 1297 (1988)  [Zh.\ Eksp.\ Teor.\ Fiz.\  {\bf 94N7}, 1 (1988)].
  
\bibitem{elisa} E. G. M. Ferreira and R. Brandenberger, 
``The Trans-Planckian Problem in the Healthy Extension of Horava-Lifshitz Gravity,''
Phys. Rev. D. {\bf 86}, 043514 (2012) [hep-th1204.5239]. 

\bibitem{yfu} J. Liu, Y. Cai and H. Li, 
``Evidences for bouncing evolution before inflation in cosmological surveys,'' 
J. Theor. Phys. {\bf 1}, 01 (2012) [astro-ph/1009.3372].
  
\bibitem{SGNG}
B.~Chen, Y.~Wang, W.~Xue and R.~Brandenberger,
  ``String Gas Cosmology and Non-Gaussianities,''
  arXiv:0712.2477 [hep-th].
  
\bibitem{YiWang}
Y.~Wang and W.~Xue,
  ``Inflation and Alternatives with Blue Tensor Spectra,''
  JCAP {\bf 1410}, no. 10, 075 (2014)
  [arXiv:1403.5817 [astro-ph.CO]].
  
\end{thebibliography}
\end{document}